\documentclass[12pt,a4paper]{article}
\usepackage{amssymb}
\usepackage{amsfonts}
\usepackage{amsmath}
\usepackage{float}
\usepackage{latexsym}
\usepackage{cite}
\begin{document}

\title{{\bf Perturbations against a Q-ball. II.\\ Contribution of nonoscillation modes}}

\author{
Mikhail~N.~Smolyakov$^{1,2}$
\\
$^1${\small{\em Skobeltsyn Institute of Nuclear Physics, Lomonosov Moscow
State University,
}}\\
{\small{\em Moscow 119991, Russia}}\\
$^2${\small{\em
Institute for Nuclear Research of the Russian Academy
of Sciences,}}\\
{\small{\em 60th October Anniversary prospect 7a, Moscow 117312,
Russia
}}}

\date{}
\maketitle

\begin{abstract}
In the present paper, discussion of perturbations against a Q-ball solution is continued. It is shown that in order to correctly describe perturbations containing nonoscillation modes, it is also necessary to consider nonlinear equations of motion for the perturbations, like in the case of oscillation modes only. It is also shown that the additivity of the charge and the energy of different modes holds for the most general nonlinear perturbation consisting of oscillation and nonoscillation modes.
\end{abstract}

\section{Introduction}
Recently, it was shown \cite{Smolyakov:2017axd} that in order to correctly describe perturbations against a background Q-ball solution \cite{Coleman:1985ki},\footnote{It should be noted that solutions of the Q-ball type were discussed in the literature earlier \cite{Finkelstein:1951zz,RosenRosenstock,Glasko,Rosen0}.} it is not sufficient to consider only the linearized equations of motion for the perturbations. In particular, the correct nonzero values of the charge and the energy of a perturbation can be obtained using the equations of motion for perturbations that contain at least the second order in the perturbations (the use of only the linearized equations of motion leads to a nonconservation of charge and energy of the perturbations). In addition, the use of such nonlinear equations also results in the additivity of the charge and the energy of different nonlinear modes forming the perturbation.

Meanwhile, only the nonlinear oscillation modes were thoroughly examined in \cite{Smolyakov:2017axd}, whereas the validity of the additivity property for the nonlinear nonoscillation modes was not checked. In the present paper I show that the use of nonlinear equations of motion for the perturbations (at least up to the second order in the perturbations) is necessary for describing both the oscillation and nonoscillation modes and for the validity of the additivity property for these nonlinear modes.

To begin with, let us briefly describe the results obtained in \cite{Smolyakov:2017axd}. The action for a complex scalar field $\phi$ in the flat $(d+1)$-dimensional space-time with $d\ge 1$, providing a Q-ball solution, has the standard form
\begin{equation}\label{action}
S=\int\left(\dot\phi^*\dot\phi-\sum_{j=1}^{d}\partial_{j}\phi^*\partial_{j}\phi-V(\phi^*\phi)\right)dtd^{d}x,
\end{equation}
where $\dot\phi=\partial_{t}\phi$. The ansatz for a Q-ball is
\begin{equation}\label{qballsolution}
\phi_{0}(t,\vec x)=e^{i\omega t}f(r).
\end{equation}
Here $r=\sqrt{{\vec x}^{2}}$ and $f(r)$ is a real function without nodes (without loss of generality, one can set $f(r)>0$ for any $r$), satisfying the boundary conditions $\partial_{r}f(r)|_{r=0}=0$, $\lim\limits_{r\to\infty}f(r)=0$ and the equation
\begin{equation}\label{eqqball}
\omega^{2}f+\Delta f-\frac{dV}{d(\phi^{*}\phi)}\biggl|_{\phi^{*}\phi=f^{2}}f=0,
\end{equation}
where $\Delta=\sum\limits_{j=1}^{d}\partial_{j}^{2}$.

The global U(1) charge is
\begin{equation}
Q=i\int\left(\phi\dot\phi^*-\phi^*\dot\phi\right)d^{d}x,
\end{equation}
giving for the Q-ball
\begin{equation}\label{qballcharge}
Q_{0}=2\omega\int f^{2}d^{d}x.
\end{equation}
In what follows, I consider only the classically stable Q-balls such that $\frac{dQ_{0}}{d\omega}<0$ \cite{Friedberg:1976me,Kumar:1979sq,GSS,Grillakis,LeePang} (for the proof of the classical stability criterion based on the use of only the linearized equations of motion for perturbations, like in \cite{VK,Kolokolov1,Kolokolov2} for the nonlinear Schr\"{o}dinger equation, see \cite{Panin:2016ooo}). The energy of the system is
\begin{equation}
E=\int\left(\dot\phi^*\dot\phi+\sum_{j=1}^{d}\partial_{j}\phi^*\partial_{j}\phi+V(\phi^*\phi)\right)d^{d}x.
\end{equation}

Now let us consider small perturbations against Q-ball solution \eqref{qballsolution}, in the form
\begin{equation}
\phi(t,\vec x)=e^{i\omega t}\bigl(f(r)+\varphi(t,\vec x)\bigr).
\end{equation}
It was shown in \cite{Smolyakov:2017axd} that in order to correctly describe the properties of perturbations consisting of oscillation modes (in particular, to obtain correct nonzero values of their charge and energy), it is necessary to consider a {\em nonlinear} equation of motion for the perturbations, at least up to the second order in the perturbations. Namely, it is necessary to consider the equation of motion
\begin{equation}\label{perteqsecond}
\omega^{2}\varphi-2i\omega\dot\varphi-\ddot\varphi+\Delta\varphi-U\varphi-S(\varphi+\varphi^{*})-
S\frac{1}{f}(\varphi^{2}+2\varphi^{*}\varphi)-J(\varphi+\varphi^{*})^{2}=0,
\end{equation}
where
\begin{equation}
U(r)=\frac{dV}{d(\phi^{*}\phi)}\biggl|_{\phi^{*}\phi=f^{2}(r)},\quad
S(r)=\frac{d^{2}V}{d(\phi^{*}\phi)^{2}}\biggl|_{\phi^{*}\phi=f^{2}(r)}f^{2}(r),\quad
J(r)=\frac{1}{2}\frac{d^{3}V}{d(\phi^{*}\phi)^{3}}\biggl|_{\phi^{*}\phi=f^{2}(r)}f^{3}(r).
\end{equation}
Equation \eqref{perteqsecond} follows directly from the equation of motion for the scalar field, which comes from action \eqref{action}. In this case, a single nonlinear mode takes the form
\begin{equation}\label{substgenintro}
\varphi_{n}(t,\vec x)=\alpha\psi_{1,n}(t,\vec x)+\alpha^{2}\psi_{2,n}(t,\vec x),
\end{equation}
where the real parameter $\alpha\ll 1$ is introduced for convenience and can be considered as the expansion parameter (in such a case, the functions $\psi_{1,n}(t,\vec x)$ and $\psi_{2,n}(t,\vec x)$ can be considered to be of the order of $f(r)$). Here $\psi_{1,n}(t,\vec x)$ satisfies the linearized equation of motion, following from \eqref{perteqsecond}. The standard ansatz for $\psi_{1,n}(t,\vec x)$ has the form \cite{Anderson:1970et,MarcVent}
\begin{equation}\label{substgeneral}
\psi_{1,n}(t,\vec x)=a_{n}(\vec x)e^{i\gamma_{n} t}+b_{n}(\vec x)e^{-i\gamma_{n}^{*} t},
\end{equation}
with the subscript $n$ enumerating different modes.\footnote{For simplicity, I take only the modes from the discrete part of the spectrum. But it is clear that the modes from the continuous part of the spectrum, if they exist, can easily be taken into account. The simplest way to do it is to put the system into a ``box'' of a finite size.} Note that $\gamma_{n}$ can be either real or purely imaginary \cite{Panin:2016ooo}. The second case corresponds to the instability mode, which is absent in the case of a classically stable Q-ball.\footnote{Since the exponentially growing instability mode destroys a classically unstable Q-ball with $\frac{dQ_{0}}{d\omega}>0$, practically it does not make sense to examine perturbations against such a Q-ball in detail.} Thus, here we can take $\gamma_{n}$ to be real (without loss of generality, $\gamma_{n}$ can be chosen, say, to be positive). The function $\psi_{2,n}(t,\vec x)$ is a correction coming from \eqref{perteqsecond}. Thus, the whole mode \eqref{substgenintro} solves nonlinear equation \eqref{perteqsecond} up to terms quadratic in the expansion parameter $\alpha$.

The charge and the energy of the perturbation up to the quadratic order in $\varphi$ are
\begin{align}\label{chargegeneral}
&Q_{p}=Q-Q_{0}=\int\Bigl(2\omega f(\varphi+\varphi^{*})+if(\dot\varphi^{*}-\dot\varphi)+2\omega\varphi^{*}\varphi+i(\dot\varphi^{*}\varphi-\varphi^{*}\dot\varphi)\Bigr)d^{d}x,\\
\label{energygeneral}
&E_{p}=E-E_{0}=\omega Q_{p}+\int\left(i\omega(\dot\varphi^{*}\varphi-\varphi^{*}\dot\varphi)+\dot\varphi^{*}\dot\varphi-\frac{1}{2}\ddot\varphi^{*}\varphi-
\frac{1}{2}\varphi^{*}\ddot\varphi\right)d^{d}x,
\end{align}
where $E_{0}$ is the Q-ball energy. Substituting
\begin{equation}\label{nonlinsolosc}
\varphi(t,\vec x)=\sum\limits_{n}\left(\alpha\psi_{1,n}(t,\vec x)+\alpha^{2}\psi_{2,n}(t,\vec x)\right)+\sum\limits_{\substack{n,n'\\n'<n}}\alpha^{2}\psi^{\times}_{n,n'}(t,\vec x),
\end{equation}
which is a solution of Eq.~\eqref{perteqsecond} (here $\psi^{\times}_{n,n'}(t,\vec x)$ stand for the overlap terms between different modes due to nonlinearity of Eq.~\eqref{perteqsecond}; see \cite{Smolyakov:2017axd} for the explicit form of the solution), into \eqref{chargegeneral} and \eqref{energygeneral}, and keeping the terms up to the second order in $\alpha$, one can get the following result \cite{Smolyakov:2017axd}:
\begin{equation}\label{chargesuper1}
Q_{p}
=\alpha^{2}\sum_{n}\int\left(\omega(\xi_{n}^{*}\xi_{n}+\eta_{n}^{*}\eta_{n})+\gamma_{n}(\xi_{n}^{*}\eta_{n}+\eta_{n}^{*}\xi_{n})-
\frac{1}{f}\frac{\partial f}{\partial\omega}S(3\xi_{n}^{*}\xi_{n}+\eta_{n}^{*}\eta_{n})-4\frac{\partial f}{\partial\omega}J\xi_{n}^{*}\xi_{n}\right)d^{d}x,
\end{equation}
\begin{equation}\label{energysuper1}
E_{p}=\omega Q_{p}+\alpha^{2}\sum_{n}\int \Bigl(\gamma_{n}^{2}(\xi_{n}^{*}\xi_{n}+\eta_{n}^{*}\eta_{n})+\omega\gamma_{n}(\xi_{n}^{*}\eta_{n}+\eta_{n}^{*}\xi_{n})\Bigr)d^{d}x,
\end{equation}
where $\xi_{n}(\vec x)=a_{n}(\vec x)+b^{*}_{n}(\vec x)$ and $\eta_{n}(\vec x)=a_{n}(\vec x)-b^{*}_{n}(\vec x)$. Formulas \eqref{chargesuper1} and \eqref{energysuper1} demonstrate the following:
\begin{enumerate}
\item Each nonlinear oscillating mode can be characterized only by solution \eqref{substgeneral} to linearized equations of motion, because $Q_{p}$ and $E_{p}$ can be represented only through the functions $\xi_{n}$, $\eta_{n}$ and the frequencies $\gamma_{n}$; see \eqref{chargesuper1} and \eqref{energysuper1}.
\item Herewith, contribution of the nonlinear part of the solution is important: it was shown in \cite{Smolyakov:2017axd} that the use of only the linear part of the perturbation for calculating its charge and energy up to the quadratic order in $\alpha$ leads to a nonconservation of charge and energy. One can also see that the function $J(r)$ in the last term of \eqref{chargesuper1} originates from the nonlinear part of Eq. \eqref{perteqsecond}.
\item The additivity property is valid for the charge and the energy of the perturbation consisting of oscillation modes; i.e., the total charge and the total energy of the perturbation are just the sums of the charges and the energies of each nonlinear oscillation mode, though the nonlinear solution for the perturbation contains explicit overlap terms between different oscillation modes.
\end{enumerate}
Thus, taking into account the nonlinear corrections $\psi_{2,n}(t,\vec x)$ and $\psi^{\times}_{n,n'}(t,\vec x)$ not only restores the conservation of charge and energy, but also leads to their additivity for the oscillation modes.

\section{Nonoscillation modes}
In \cite{Smolyakov:2017axd}, the additivity property was proven to be valid for the oscillation modes only, whose linear parts have the form \eqref{substgeneral}. However, the linearized equation of motion
\begin{equation}\label{lineqgen}
\omega^{2}\varphi_{lin}-2i\omega\dot\varphi_{lin}-\ddot\varphi_{lin}+\Delta\varphi_{lin}-U\varphi_{lin}-S(\varphi_{lin}+\varphi^{*}_{lin})=0
\end{equation}
provides other solutions, which have completely different forms. In particular, one can check that the modes (in fact, all these modes correspond to different symmetries of the system)
\begin{align}\label{modeU1}
&\varphi_{lin}^{U(1)}(t,\vec x)=\epsilon if(r),\\
\label{modeTrans}
&\varphi_{lin}^{Tr,j}(t,\vec x)=l_{j}\,\partial_{j}f(r),\\
\label{modeLorentz}
&\varphi_{lin}^{Lor,j}(t,\vec x)=v_{j}\left(t\partial_{j}f(r)+i\omega x_{j}f(r)\right),\\
\label{modeCharge}
&\varphi_{lin}^{\omega}(t,\vec x)=\delta\omega\left(\frac{\partial f(r)}{\partial\omega}+itf(r)\right)
\end{align}
satisfy Eq. \eqref{lineqgen}. Formally, solutions \eqref{modeLorentz} and \eqref{modeCharge} can be used for $t\lesssim \frac{1}{\omega}$. Here $\epsilon$, $l_{j}$, $v_{j}$, and $\delta\omega$ are the corresponding small (expansion) parameters and $x_{j}=x^{j}$, $j=1,...,d$. Mode \eqref{modeU1} corresponds to the global $U(1)$ symmetry, mode \eqref{modeTrans} corresponds to the translational symmetry, mode \eqref{modeLorentz} corresponds to the Lorentz symmetry, and mode \eqref{modeCharge} corresponds to the change of the Q-ball frequency $\omega$. Solutions \eqref{modeU1}--\eqref{modeLorentz} were discussed in \cite{Smolyakov:2017axd} separately; it was shown that the nonlinear corrections are important for obtaining correct values of $Q_{p}$ and $E_{p}$ for these modes. Namely, it was shown that up to the quadratic order in the expansion parameters
\begin{align}\label{QEU1}
&Q_{p}^{U(1)}=0,\qquad E_{p}^{U(1)}=0,\hspace{1.8cm} \varphi^{U(1)}=\epsilon if-\epsilon^{2}\frac{1}{2}f,
\end{align}
\begin{align}
\label{QETrj}
&Q_{p}^{Tr,j}=0,\qquad E_{p}^{Tr,j}=0,\hspace{1.9cm} \varphi^{Tr,j}=l_{j}\,\partial_{j}f+l_{j}^{2}\frac{1}{2}\partial_{j}^{2}f,\\ \nonumber
&Q_{p}^{Lor,j}=0,\qquad E_{p}^{Lor,j}=\frac{E_{0}v_{j}^{2}}{2},\hspace{0.9cm} \varphi^{Lor,j}=v_{j}\left(t\partial_{j}f+i\omega x_{j}f\right)\\ \label{QELorj}&\hspace{4cm}+v_{j}^{2}\left(\frac{1}{2}x_{j}\partial_{j}f+\frac{1}{2}t^{2}\partial_{j}^{2}f-\frac{1}{2}\omega^{2}x_{j}^{2}f+i\left(\omega tx_{j}\partial_{j}f+\frac{1}{2}\omega t f\right)\right).
\end{align}
As a trivial demonstration of the necessity of the nonlinear parts of the solutions, one can consider the case of the mode corresponding to the global $U(1)$ symmetry. Using \eqref{chargegeneral} and \eqref{energygeneral}, it is easy to check that the use of only the linear part of the solution, i.e., $\varphi^{U(1)}=\varphi^{U(1)}_{lin}=\epsilon if$, leads to \cite{Smolyakov:2017axd}
\begin{equation}
Q_{p}^{U(1)}=2\epsilon^{2}\omega\int f^{2}d^{d}x,\qquad E_{p}^{U(1)}=2\epsilon^{2}\omega^{2}\int f^{2}d^{d}x,
\end{equation}
which looks unphysical, because multiplication of the initial solution $f$ by $e^{i\epsilon}$ does not change the total charge and energy, so we expect to get $Q_{p}^{U(1)}=0$, $E_{p}^{U(1)}=0$. However, we indeed get $Q_{p}^{U(1)}=0$, $E_{p}^{U(1)}=0$ (of course, up to the terms $\sim\epsilon^{2}$ in this approximation) for the nonlinear solution presented in \eqref{QEU1}. Analogous reasonings can be applied to other modes. Thus, the values of $Q_{p}$ and $E_{p}$ in \eqref{QEU1}--\eqref{QELorj} are expected from the physical point of view. However, these modes were considered in \cite{Smolyakov:2017axd} separately, without taking into account that the whole perturbation may include all these modes and, in addition, the oscillation modes.

It is not clear whether the additivity property is valid for the nonoscillation nonlinear modes in the general case. In principle, the additivity property is expected to hold for the mode corresponding to the global $U(1)$ symmetry \eqref{QEU1} and for the translational mode \eqref{QETrj}. Indeed, since multiplication of the whole solution by a phase factor or its translation in a spatial coordinate does not change the total charge and energy of the system, there should not be overlap terms between these modes and other nonlinear modes of the perturbation in the expressions for the charge and the energy (but there {\em should be} overlap terms in the nonlinear solution for the perturbation). Meanwhile, since mode \eqref{QELorj} has a nonzero contribution to the energy, whereas the mode corresponding to the change of the Q-ball frequency changes the charge and the energy of the system (it is demonstrated below), the answer is not so obvious. Moreover, naively one may suppose that the additivity property for the nonlinear oscillation modes is valid because of their oscillation form, which is the same for each mode. Below I demonstrate that the additivity property (of course, up to the terms quadratic in the expansion parameters) is valid for the nonlinear perturbation consisting of the oscillation modes and of the nonlinear modes corresponding to \eqref{modeU1}--\eqref{modeCharge}, too.

\subsection{The mode corresponding to the change of the Q-ball frequency}
To start with, let us first consider the nonlinear mode, corresponding to \eqref{modeCharge}, separately. One can check that
\begin{equation}\label{modeOmeganonlin}
\varphi^{\omega}(t,\vec x)=\delta\omega\left(\frac{\partial f}{\partial\omega}+itf\right)+\delta\omega^{2}\left(\frac{1}{2}\frac{\partial^{2}f}{\partial\omega^{2}}+it\frac{\partial f}{\partial\omega}-\frac{1}{2}t^{2}f\right)
\end{equation}
satisfies Eq.~\eqref{perteqsecond} up to the terms $\sim \delta\omega^{2}$. When checking the fulfillment of Eq.~\eqref{perteqsecond}, it is convenient to use the equations
\begin{align}\label{eqsL1}
&\hat L_{2}f=0,\\ \label{eqsL2}
&\partial_{\omega}(\hat L_{2}f)\equiv \hat L_{1}\partial_{\omega}f-2\omega f=0,\\ \label{eqsL3}
&\partial^{2}_{\omega}(\hat L_{2}f)\equiv \hat L_{1}\partial^{2}_{\omega}f-4\omega\partial_{\omega}f-2f+(6Sf^{-1}+8J)(\partial_{\omega}f)^{2}=0,
\end{align}
where the operators $\hat L_{1}$ and $\hat L_{2}$ are defined as
\begin{align}\label{L1eq}
&\hat L_{1}=-\Delta+U(r)+2S(r)-\omega^{2},\\ \label{L2eq}
&\hat L_{2}=-\Delta+U(r)-\omega^{2}.
\end{align}
Equation \eqref{eqsL1} is just Eq.~\eqref{eqqball} for the Q-ball profile.

Note that the terms $\sim t$ and $\sim t^{2}$ in \eqref{modeOmeganonlin}, which grow with time, do not indicate any instability. Indeed, these terms are just the parts of the first two terms of the decomposition of $e^{i(\omega+\delta\omega)t}f(r,\omega+\delta\omega)-e^{i\omega t}f(r,\omega)$, and it is clear that the latter function is bounded in time.

Substituting \eqref{modeOmeganonlin} into \eqref{chargegeneral} and retaining the terms $\sim \delta\omega$ and $\sim \delta\omega^{2}$, after some straightforward calculations one gets
\begin{equation}\label{Qomega}
Q_{p}^{\omega}=\frac{dQ_{0}}{d\omega}\delta\omega+\frac{1}{2}\frac{d^{2}Q_{0}}{d\omega^{2}}\delta\omega^{2},
\end{equation}
which is the expected result. For the energy, we get
\begin{equation}\label{Eomega}
E_{p}^{\omega}=\omega Q_{p}^{\omega}+\frac{1}{2}\frac{dQ_{0}}{d\omega}\delta\omega^{2}
=\omega\frac{dQ_{0}}{d\omega}\delta\omega+\frac{1}{2}\left(\omega\frac{d^{2}Q_{0}}{d\omega^{2}}+\frac{dQ_{0}}{d\omega}\right)\delta\omega^{2}.
\end{equation}
An important point is that $\varphi^{\omega}$ is the only mode providing nonzero contributions to the charge and the energy, which are linear in the small parameter (i.e., the terms $\sim\delta\omega$ in \eqref{Qomega} and \eqref{Eomega}). Indeed, all the other modes have no nonzero contributions $\sim\alpha$, $\sim\epsilon$, $\sim l_{j}$ or $\sim v_{j}$ to $Q_{p}$ and $E_{p}$; see \eqref{chargesuper1}, \eqref{energysuper1}, and \eqref{QEU1}--\eqref{QELorj}. We discuss this property of the $\varphi^{\omega}$ mode later.

There is an additional way to check the validity of formula \eqref{Eomega}. It is well known that the relation
\begin{equation}\label{dEdQ}
\frac{dE_{0}}{d\omega}=\omega\frac{dQ_{0}}{d\omega}
\end{equation}
holds for Q-balls. Differentiating this relation with respect to $\omega$, one gets
\begin{equation}\label{d2EdQ2}
\frac{d^{2}E_{0}}{d\omega^{2}}=\omega\frac{d^{2}Q_{0}}{d\omega^{2}}+\frac{dQ_{0}}{d\omega}.
\end{equation}
Substituting \eqref{dEdQ} and \eqref{d2EdQ2} into \eqref{Eomega}, we get the expected result
\begin{equation}\label{Eomega1}
E_{p}^{\omega}=\frac{dE_{0}}{d\omega}\delta\omega+\frac{1}{2}\frac{d^{2}E_{0}}{d\omega^{2}}\delta\omega^{2}.
\end{equation}

\subsection{Structure of the nonlinear perturbation}
Suppose that we have a perturbation whose linear part has the most general form
\begin{equation}\label{linpertfull}
\varphi_{lin}(t,\vec x)=\sum\limits_{n}\alpha\psi_{1,n}(t,\vec x)+\sum\limits_{m}\beta_{m}\theta_{1,m}(t,\vec x),
\end{equation}
where $\theta_{1,m}(t,\vec x)$ corresponds to solutions \eqref{modeU1}--\eqref{modeCharge} of the linearized equation of motion and $\beta_{m}$ denote the expansion parameters $\epsilon$, $l_{j}$, $v_{j}$, and $\delta\omega$ (with $j=1,...,d$). For the convenience, these parameters are presented in  \eqref{linpertfull} explicitly. In this case, the full nonlinear perturbation up to the terms quadratic in $\alpha$ and $\beta_{m}$ takes the form (see also \eqref{nonlinsolosc})
\begin{align}\nonumber
\varphi(t,\vec x)=\sum\limits_{n}\alpha\psi_{1,n}(t,\vec x)+\sum\limits_{n}\alpha^{2}\psi_{2,n}(t,\vec x)+\sum\limits_{\substack{n,n'\\n'<n}}\alpha^{2}\psi^{\times}_{n,n'}(t,\vec x)\\ \nonumber
+\sum\limits_{m}\beta_{m}\theta_{1,m}(t,\vec x)+\sum\limits_{m}\beta_{m}^{2}\theta_{2,m}(t,\vec x)+
\sum\limits_{\substack{m,m'\\m'<m}}\beta_{m}\beta_{m'}\theta^{\times}_{m,m'}(t,\vec x)\\+
\sum\limits_{\substack{n,m}}\alpha\beta_{m}\kappa^{\times}_{n,m}(t,\vec x).
\end{align}
Here the term $\sim\theta_{2,m}(t,\vec x)$ corresponds to the nonlinear part of the $m$th single nonoscillation mode; the term $\sim\theta^{\times}_{m,m'}(t,\vec x)$ corresponds to the nonlinear overlap solution between $m$th and $m'$th nonoscillation modes; and the term $\sim\kappa^{\times}_{n,m}$ corresponds to the nonlinear overlap solution between the $n$th oscillation mode and $m$th nonoscillation mode. It was shown in \cite{Smolyakov:2017axd} that the terms in $Q_{p}$ and $E_{p}$ containing $\psi^{\times}_{n,n'}(t,\vec x)$, $\psi_{1,n}(t,\vec x)$, and $\psi_{1,n'}(t,\vec x)$ that are $\sim\alpha^{2}$, i.e., the overlap terms between different oscillation modes, compensate each other in $Q_{p}$ and $E_{p}$. Below we calculate the contributions $\sim\alpha\beta_{m}$ and $\sim\beta_{m}\beta_{m'}$ to $Q_{p}$ and $E_{p}$.

\subsection{Overlap terms between oscillation and nonoscillation modes}
First, let us consider the case of one oscillation mode and one nonoscillation mode. One can see that the linear part of each nonoscillation mode \eqref{modeU1}--\eqref{modeCharge} can be rewritten in a universal form,
\begin{equation}\label{nonoscuniv}
\theta_{1}(t,\vec x)=A(\vec x)+tB(\vec x),
\end{equation}
where $A(\vec x)$ and $B(\vec x)$ are complex functions (for simplicity, here I drop the subscript $m$ labeling the mode). Substituting this ansatz into linear equation \eqref{lineqgen}, we obtain the system of equations
\begin{align}\label{eqlinnonosc1}
&\hat L_{1}(A+A^{*})=-2i\omega(B-B^{*}),\\ \label{eqlinnonosc2}
&\hat L_{1}(B+B^{*})=0,\\ \label{eqlinnonosc3}
&\hat L_{2}(A-A^{*})=-2i\omega(B+B^{*}),\\ \label{eqlinnonosc4}
&\hat L_{2}(B-B^{*})=0,
\end{align}
where the operators $\hat L_{1}$ and $\hat L_{2}$ are defined by \eqref{L1eq} and \eqref{L2eq}.

The linear part of any oscillation mode is defined by
\begin{equation}\label{oscuniv}
\psi_{1}(t,\vec x)=a(\vec x)e^{i\gamma t}+b(\vec x)e^{-i\gamma t},
\end{equation}
with $\gamma\neq 0$; see \eqref{substgeneral} (again, for simplicity, here I drop the subscript $n$). The linearized equations of motion can be rewritten in the useful form \cite{Smolyakov:2017axd}
\begin{align}\label{lineq1intro}
\hat L_{1}\xi-2\omega\gamma\eta-\gamma^{2}\xi=0,\\ \label{lineq2intro}
\hat L_{2}\eta-2\omega\gamma\xi-\gamma^{2}\eta=0,
\end{align}
where $\xi(\vec x)=a(\vec x)+b^{*}(\vec x)$ and $\eta(\vec x)=a(\vec x)-b^{*}(\vec x)$.

The structure of solutions \eqref{nonoscuniv} and \eqref{oscuniv}, together with the form of nonlinear equation of motion \eqref{perteqsecond}, suggests that the nonlinear overlap term between these modes $\kappa^{\times}(t,\vec x)$ has the form
\begin{align}\label{overlap1}
\kappa^{\times}(t,\vec x)+\left(\kappa^{\times}(t,\vec x)\right)^{*}=\rho_{1}(\vec x)e^{i\gamma t}+\rho_{1}^{*}(\vec x)e^{-i\gamma t}+t\left(\rho_{2}(\vec x)e^{i\gamma t}+\rho_{2}^{*}(\vec x)e^{-i\gamma t}\right),\\ \label{overlap2}
\kappa^{\times}(t,\vec x)-\left(\kappa^{\times}(t,\vec x)\right)^{*}=\delta_{1}(\vec x)e^{i\gamma t}-\delta_{1}^{*}(\vec x)e^{-i\gamma t}+t\left(\delta_{2}(\vec x)e^{i\gamma t}-\delta_{2}^{*}(\vec x)e^{-i\gamma t}\right).
\end{align}
Substituting \eqref{nonoscuniv}, \eqref{oscuniv}, \eqref{overlap1}, and \eqref{overlap2} into nonlinear equation of motion \eqref{perteqsecond}, keeping the corresponding overlap terms and isolating the terms with different dependence on time $t$, we can get the equations of motion for the functions $\rho_{1}(\vec x)$, $\rho_{2}(\vec x)$, $\delta_{1}(\vec x)$, and $\delta_{2}(\vec x)$,
\begin{align}
&-\gamma^{2}\rho_{1}+2i\gamma\rho_{2}-2\omega\gamma\delta_{1}+2i\omega\delta_{2}+\hat L_{1}\rho_{1}=-\left(3\frac{S}{f}+4J\right)\xi(A+A^{*})+\frac{S}{f}\eta(A-A^{*}),\\
&-\gamma^{2}\rho_{2}-2\omega\gamma\delta_{2}+\hat L_{1}\rho_{2}=-\left(3\frac{S}{f}+4J\right)\xi(B+B^{*})+\frac{S}{f}\eta(B-B^{*}),\\ \label{overeq3}
&-\gamma^{2}\delta_{1}+2i\gamma\delta_{2}-2\omega\gamma\rho_{1}+2i\omega\rho_{2}+\hat L_{2}\delta_{1}=-\frac{S}{f}\xi(A-A^{*})-\frac{S}{f}\eta(A+A^{*}),\\ \label{overeq4}
&-\gamma^{2}\delta_{2}-2\omega\gamma\rho_{2}+\hat L_{2}\delta_{2}=-\frac{S}{f}\xi(B-B^{*})-\frac{S}{f}\eta(B+B^{*}).
\end{align}
Not all these equations are used in the subsequent calculations, only Eqs.~\eqref{overeq3} and \eqref{overeq4}.

Now let us substitute \eqref{nonoscuniv}, \eqref{oscuniv} and \eqref{overlap1}, \eqref{overlap2} (multiplied by the corresponding small parameters) into \eqref{chargegeneral}. After some algebra, we get for the overlap term
\begin{align}\nonumber
Q_{p}^{\times}=\alpha\beta\int d^{d}x\biggl(e^{i\gamma t}\biggl[2\omega f\rho_{1}+\gamma f\delta_{1}-if\delta_{2}\\ \nonumber
+\frac{1}{2}\Bigl((A+A^{*})(2\omega\xi+\gamma\eta)-(A-A^{*})(2\omega\eta+\gamma\xi)-i(B-B^{*})\xi+i(B+B^{*})\eta\Bigr)\biggr]\\ \label{chargeoverlap}
+te^{i\gamma t}\biggl[2\omega f\rho_{2}+\gamma f\delta_{2}+\frac{1}{2}\Bigl((B+B^{*})(2\omega\xi+\gamma\eta)
-(B-B^{*})(2\omega\eta+\gamma\xi)\Bigr)\biggr]+\textrm{c.c.}\biggr).
\end{align}

To calculate $Q_{p}^{\times}$, let us perform the following steps. First, let us take Eq.~\eqref{overeq4}, multiply it by $f$, and integrate the result over the space. We get
\begin{equation}\label{intrelaux1}
\int d^{d}x\left(2\omega f\rho_{2}+\gamma f\delta_{2}\right)=\frac{1}{\gamma}\int d^{d}x\left(S\xi(B-B^{*})+S\eta(B+B^{*})\right),
\end{equation}
where we have used the fact that $\hat L_{2}f=0$ (see Eq.~\eqref{eqsL1}). Analogously, let us take Eq.~\eqref{overeq3}, multiply it by $f$, and integrate the result over the space. Using $\hat L_{2}f=0$ and relation \eqref{intrelaux1}, we get
\begin{align}\nonumber
&\int d^{d}x\left(2\omega f\rho_{1}+\gamma f\delta_{1}-if\delta_{2}\right)\\ \label{intrelaux2}
&=\frac{1}{\gamma}\int d^{d}x\left(S\xi(A-A^{*})+S\eta(A+A^{*})\right)+\frac{i}{\gamma^{2}}\int d^{d}x\left(S\xi(B-B^{*})+S\eta(B+B^{*})\right).
\end{align}

Second, let us take the integral
\begin{equation}
\int d^{d}x(B-B^{*})\hat L_{1}\xi.
\end{equation}
On the one hand,
\begin{align}\nonumber
&\int d^{d}x(B-B^{*})\hat L_{1}\xi=\int d^{d}x(B-B^{*})\left(\hat L_{2}+2S\right)\xi\\ \label{intcomb1-1}
&=\int d^{d}x\,\xi\hat L_{2}(B-B^{*})+2\int d^{d}x(B-B^{*})S\xi=2\int d^{d}x(B-B^{*})S\xi,
\end{align}
where we have used definition \eqref{L1eq} of the operator $\hat L_{1}$ and Eq.~\eqref{eqlinnonosc4}. On the other hand,
\begin{equation}\label{intcomb1-2}
\int d^{d}x(B-B^{*})\hat L_{1}\xi=\int d^{d}x(B-B^{*})(2\omega\gamma\eta+\gamma^{2}\xi),
\end{equation}
where we have used \eqref{lineq1intro}. Combining relations \eqref{intcomb1-1} and \eqref{intcomb1-2}, we obtain
\begin{equation}\label{intcomb1}
\int d^{d}x S\xi(B-B^{*})=\frac{1}{2}\int d^{d}x(B-B^{*})(2\omega\gamma\eta+\gamma^{2}\xi).
\end{equation}
Analogously, using the integral
\begin{equation}
\int d^{d}x(B+B^{*})\hat L_{2}\eta
\end{equation}
and Eqs.~\eqref{eqlinnonosc2} and \eqref{lineq2intro}, one can get
\begin{equation}\label{intcomb2}
\int d^{d}x S\eta(B+B^{*})=-\frac{1}{2}\int d^{d}x(B+B^{*})(2\omega\gamma\xi+\gamma^{2}\eta);
\end{equation}
using the integral
\begin{equation}
\int d^{d}x(A-A^{*})\hat L_{1}\xi
\end{equation}
and Eqs.~\eqref{eqlinnonosc3} and \eqref{lineq1intro}, one can get
\begin{equation}\label{intcomb3}
\int d^{d}x S\xi(A-A^{*})=\frac{1}{2}\int d^{d}x(A-A^{*})(2\omega\gamma\eta+\gamma^{2}\xi)+i\int d^{d}x\,\omega\xi(B+B^{*});
\end{equation}
using the integral
\begin{equation}
\int d^{d}x(A+A^{*})\hat L_{2}\eta
\end{equation}
and Eqs.~\eqref{eqlinnonosc1} and \eqref{lineq2intro}, one can get
\begin{equation}\label{intcomb4}
\int d^{d}x S\eta(A+A^{*})=-\frac{1}{2}\int d^{d}x(A+A^{*})(2\omega\gamma\xi+\gamma^{2}\eta)-i\int d^{d}x\,\omega\eta(B-B^{*}).
\end{equation}

And third, combining relations \eqref{intrelaux1} and \eqref{intrelaux2} and relations \eqref{intcomb1}, \eqref{intcomb2}, \eqref{intcomb3}, and \eqref{intcomb4}, we get
\begin{equation}\label{intrelQfinal1}
\int d^{d}x\left(2\omega f\rho_{2}+\gamma f\delta_{2}\right)
=\frac{1}{2}\int d^{d}x\Bigl((B-B^{*})(2\omega\eta+\gamma\xi)-(B+B^{*})(2\omega\xi+\gamma\eta)\Bigr)
\end{equation}
and
\begin{align}\nonumber
&\int d^{d}x\left(2\omega f\rho_{1}+\gamma f\delta_{1}-if\delta_{2}\right)\\ \label{intrelQfinal2}
&=\frac{1}{2}\int d^{d}x
\Bigl((A-A^{*})(2\omega\eta+\gamma\xi)-(A+A^{*})(2\omega\xi+\gamma\eta)+i\xi(B-B^{*})-i\eta(B+B^{*})\Bigr).
\end{align}

Finally, substituting relations \eqref{intrelQfinal1} and \eqref{intrelQfinal2} into \eqref{chargeoverlap}, we easily get
\begin{equation}\label{zerooverlapQ}
Q_{p}^{\times}=0.
\end{equation}

Now let us turn to the calculation of $E_{p}^{\times}$. According to the definition of the perturbation energy \eqref{energygeneral} and taking into account \eqref{zerooverlapQ}, it is clear that only the linear part of the perturbation is necessary for calculating $E_{p}^{\times}$. Substituting \eqref{nonoscuniv} and \eqref{oscuniv} (multiplied by the corresponding small parameters) into \eqref{energygeneral}, after some algebra we get
\begin{align}\nonumber
E_{p}^{\times}=\frac{\alpha\beta}{4}\int d^{d}x\biggl(e^{i\gamma t}\Bigl[(A+A^{*})(2\omega\gamma\eta+\gamma^{2}\xi)-(A-A^{*})(2\omega\gamma\xi+\gamma^{2}\eta)\\ \nonumber+2i(B+B^{*})(\omega\eta+\gamma\xi)-2i(B-B^{*})(\omega\xi+\gamma\eta)\Bigr]\\ \label{energyoverlap}
+te^{i\gamma t}\Bigl[(B+B^{*})(2\omega\gamma\eta+\gamma^{2}\xi)-(B-B^{*})(2\omega\gamma\xi+\gamma^{2}\eta)\Bigr]+\textrm{c.c.}\biggr).
\end{align}
Note that although the terms in \eqref{energyoverlap} look similar to those in \eqref{chargeoverlap}, their structure is different.

Let us take the integral
\begin{equation}
\int d^{d}x(A+A^{*})\hat L_{1}\xi.
\end{equation}
On the one hand, from Eq.~\eqref{lineq1intro} it follows that
\begin{equation}
\int d^{d}x(A+A^{*})\hat L_{1}\xi=\int d^{d}x(A+A^{*})(2\omega\gamma\eta+\gamma^{2}\xi).
\end{equation}
On the other hand, from Eq.~\eqref{eqlinnonosc1} it follows that
\begin{equation}
\int d^{d}x(A+A^{*})\hat L_{1}\xi=\int d^{d}x\,\xi\hat L_{1}(A+A^{*})=-2i\int d^{d}x\,\omega\xi(B-B^{*}).
\end{equation}
Thus,
\begin{equation}\label{intauxenergy1}
\int d^{d}x(A+A^{*})(2\omega\gamma\eta+\gamma^{2}\xi)=-2i\int d^{d}x\,\omega\xi(B-B^{*}).
\end{equation}
Analogously, using the integral
\begin{equation}
\int d^{d}x(A-A^{*})\hat L_{2}\eta
\end{equation}
and Eqs.~\eqref{lineq2intro} and \eqref{eqlinnonosc3}, one can get
\begin{equation}\label{intauxenergy2}
\int d^{d}x(A-A^{*})(2\omega\gamma\xi+\gamma^{2}\eta)=-2i\int d^{d}x\,\omega\eta(B+B^{*}).
\end{equation}
Substituting relations \eqref{intauxenergy1} and \eqref{intauxenergy2} into \eqref{energyoverlap}, we obtain
\begin{equation}\label{energyoverlap2}
E_{p}^{\times}=\frac{\alpha\beta}{4}\left(\frac{2i}{\gamma}+t\right)e^{i\gamma t}\int d^{d}x\Bigl((B+B^{*})(2\omega\gamma\eta+\gamma^{2}\xi)-(B-B^{*})(2\omega\gamma\xi+\gamma^{2}\eta)\Bigr)+\textrm{c.c.}
\end{equation}

Now let us consider the integral
\begin{equation}
\int d^{d}x(B+B^{*})\hat L_{1}\xi.
\end{equation}
On the one hand, from Eq.~\eqref{lineq1intro} it follows that
\begin{equation}
\int d^{d}x(B+B^{*})\hat L_{1}\xi=\int d^{d}x(B+B^{*})(2\omega\gamma\eta+\gamma^{2}\xi).
\end{equation}
On the other hand, from Eq.~\eqref{eqlinnonosc2} it follows that
\begin{equation}
\int d^{d}x(B+B^{*})\hat L_{1}\xi=\int d^{d}x\,\xi\hat L_{1}(B+B^{*})=0.
\end{equation}
Thus,
\begin{equation}\label{intauxenergy3}
\int d^{d}x(B+B^{*})(2\omega\gamma\eta+\gamma^{2}\xi)=0.
\end{equation}
Analogously, using the integral
\begin{equation}
\int d^{d}x(B-B^{*})\hat L_{2}\eta
\end{equation}
and Eqs.~\eqref{lineq2intro} and \eqref{eqlinnonosc4}, one can get
\begin{equation}\label{intauxenergy4}
\int d^{d}x(B-B^{*})(2\omega\gamma\xi+\gamma^{2}\eta)=0.
\end{equation}
Substituting \eqref{intauxenergy3} and \eqref{intauxenergy4} into \eqref{energyoverlap2}, we obtain
\begin{equation}
E_{p}^{\times}=0.
\end{equation}

Thus, we see that the overlap terms in $Q_{p}$ and $E_{p}$ between any nonoscillation mode, whose linear part has the form \eqref{nonoscuniv} and satisfies Eqs.~\eqref{eqlinnonosc1}--\eqref{eqlinnonosc4}, and any oscillation mode, whose linear part has the form \eqref{oscuniv} and satisfies Eqs.~\eqref{lineq1intro}--\eqref{lineq2intro}, vanish.

\subsection{Overlap terms between different nonoscillation modes}
Now let us examine the overlap terms in $Q_{p}$ and $E_{p}$ between the nonoscillation modes themselves. It turns out that the simplest way to demonstrate that the corresponding terms also vanish is just to obtain the explicit solutions to the nonlinear equations of motion for the perturbations and to evaluate all the necessary integrals. Below the explicit formulas for the modes, whose linear parts are \eqref{modeU1}--\eqref{modeCharge}, together with the corresponding nonlinear overlap solutions, are presented.
\begin{enumerate}
\item
\begin{align}\label{nonoscoverlap1}
&\beta_{m}\theta_{1,m}=\epsilon if,\qquad \beta_{m'}\theta_{1,m'}=l_{k}\,\partial_{k}f,\\
&\beta_{m}\beta_{m'}\theta^{\times}_{m,m'}=l_{k}\epsilon\Bigl(i\partial_{k}f\Bigr).
\end{align}
\item Here $j\neq k$:
\begin{align}
&\beta_{m}\theta_{1,m}=l_{j}\,\partial_{j}f,\qquad \beta_{m'}\theta_{1,m'}=l_{k}\,\partial_{k}f,\\
&\beta_{m}\beta_{m'}\theta^{\times}_{m,m'}=l_{j}l_{k}\Bigl(\partial_{j}\partial_{k}f\Bigr).
\end{align}
\item
\begin{align}
&\beta_{m}\theta_{1,m}=v_{j}\left(t\partial_{j}f+i\omega x_{j}f\right),\qquad \beta_{m'}\theta_{1,m'}=\epsilon if,\\
&\beta_{m}\beta_{m'}\theta^{\times}_{m,m'}=v_{j}\epsilon\Bigl(it\partial_{j}f-\omega x_{j}f\Bigr).
\end{align}
\item
\begin{align}
&\beta_{m}\theta_{1,m}=\delta\omega\left(\partial_{\omega}f+itf\right),\qquad \beta_{m'}\theta_{1,m'}=\epsilon if,\\
&\beta_{m}\beta_{m'}\theta^{\times}_{m,m'}=\epsilon\delta\omega\Bigl(i\partial_{\omega}f-tf\Bigr).
\end{align}
\item
\begin{align}
&\beta_{m}\theta_{1,m}=v_{j}\left(t\partial_{j}f+i\omega x_{j}f\right),\qquad \beta_{m'}\theta_{1,m'}=l_{k}\,\partial_{k}f,\\
&\beta_{m}\beta_{m'}\theta^{\times}_{m,m'}=v_{j}l_{k}\Bigl(i\omega x_{j}\partial_{k}f+i\omega\delta_{jk}f+t\partial_{j}\partial_{k}f\Bigr),
\end{align}
where $\delta_{jk}$ is the Kronecker delta.
\item
\begin{align}
&\beta_{m}\theta_{1,m}=\delta\omega\left(\partial_{\omega}f+itf\right),\qquad \beta_{m'}\theta_{1,m'}=l_{k}\,\partial_{k}f,\\ \label{nonoscoverlap1-2}
&\beta_{m}\beta_{m'}\theta^{\times}_{m,m'}=l_{k}\delta\omega\Bigl(it\partial_{k}f+\partial_{k}\partial_{\omega}f\Bigr).
\end{align}
\item
\begin{align}\label{nonoscoverlap1-3}
&\beta_{m}\theta_{1,m}=v_{j}\left(t\partial_{j}f+i\omega x_{j}f\right),\qquad \beta_{m'}\theta_{1,m'}=\delta\omega\left(\partial_{\omega}f+itf\right),\\
&\beta_{m}\beta_{m'}\theta^{\times}_{m,m'}=v_{j}\delta\omega\Bigl(ix_{j}\left(f+\omega\partial_{\omega}f\right)+t\left(\partial_{j}\partial_{\omega}f-\omega x_{j}f\right)+it^{2}\partial_{j}f\Bigr).
\end{align}
\item Here $j\neq k$:
\begin{align}
&\beta_{m}\theta_{1,m}=v_{j}\left(t\partial_{j}f+i\omega x_{j}f\right),\qquad \beta_{m'}\theta_{1,m'}=v_{k}\left(t\partial_{k}f+i\omega x_{k}f\right),\\ \label{nonoscoverlap2}
&\beta_{m}\beta_{m'}\theta^{\times}_{m,m'}=v_{j}v_{k}
\left(\left(\frac{1}{2}+i\omega t\right)(x_{j}\partial_{k}f+x_{k}\partial_{j}f)-\omega^{2}x_{j}x_{k}f+t^{2}\partial_{j}\partial_{k}f\right).
\end{align}
\end{enumerate}
Cases 1--6 (formulas \eqref{nonoscoverlap1}--\eqref{nonoscoverlap1-2}) are rather trivial, whereas cases 7--8 (formulas \eqref{nonoscoverlap1-3}--\eqref{nonoscoverlap2}) turn out to be more involved.

Substituting each group of \eqref{nonoscoverlap1}--\eqref{nonoscoverlap2} into nonlinear equation of motion \eqref{perteqsecond} and keeping the terms $\sim \beta_{m}\beta_{m'}$, one can check that Eq.~\eqref{perteqsecond} is indeed satisfied. The corresponding calculations are straightforward, though rather bulky; during the calculations, one should use Eqs.~\eqref{eqsL1}--\eqref{eqsL3}, and the equations
\begin{align}
&\partial_{j}(\hat L_{2}f)\equiv \hat L_{1}\partial_{j}f=0,\\
&\partial_{j}\partial_{k}(\hat L_{2}f)\equiv \hat L_{1}\partial_{j}\partial_{k}f+(6Sf^{-1}+8J)\partial_{j}f\partial_{k}f=0,\\
&\partial_{\omega}\partial_{k}(\hat L_{2}f)\equiv \hat L_{1}\partial_{\omega}\partial_{k}f-2\omega\partial_{k}f+(6Sf^{-1}+8J)\partial_{k}f\partial_{\omega}f=0.
\end{align}
Then, substituting each group of \eqref{nonoscoverlap1}--\eqref{nonoscoverlap2} into \eqref{chargegeneral} and \eqref{energygeneral}, isolating the terms $\sim \beta_{m}\beta_{m'}$ and performing an integration by parts in some terms (all the calculations are straightforward, though also rather bulky), one can easily check that the overlap terms in $Q_{p}$ and $E_{p}$ between all nonoscillation modes presented above also vanish.

\section{Discussion}
As was shown in the previous sections, though the nonlinear overlap solutions between different modes are not equal to 0 (formulas \eqref{overlap1} and \eqref{overlap2} and the corresponding formulas in \eqref{nonoscoverlap1}--\eqref{nonoscoverlap2}), the overlap terms between different nonlinear modes in $Q_{p}$ and $E_{p}$, calculated up to the second order in the expansion parameters, vanish. It means that the additivity property is valid for the charge and the energy of all nonlinear modes forming the perturbation. Namely, for the perturbation such that its linear part has the form
\begin{align}\nonumber
&\varphi_{lin}(t,\vec x)=\epsilon if(r)+\sum\limits_{j=1}^{d}l_{j}\,\partial_{j}f(r)+\delta\omega\left(\partial_{\omega} f(r)+itf(r)\right)\\
&+\sum\limits_{j=1}^{d}v_{j}\left(t\partial_{j}f(r)+i\omega x_{j}f(r)\right)+\alpha\sum\limits_{n}\left(a_{n}(\vec x)e^{i\gamma_{n} t}+b_{n}(\vec x)e^{-i\gamma_{n} t}\right),
\end{align}
with $a_{n}(\vec x)$, $b_{n}(\vec x)$ and $\gamma_{n}$ satisfying Eqs.~\eqref{lineq1intro} and \eqref{lineq2intro}, its charge and energy take the form
\begin{align}\nonumber
&Q_{p}=\frac{dQ_{0}}{d\omega}\delta\omega+\frac{1}{2}\frac{d^{2}Q_{0}}{d\omega^{2}}\delta\omega^{2}\\&
+\alpha^{2}\sum_{n}\int\left(\omega(\xi_{n}^{*}\xi_{n}+\eta_{n}^{*}\eta_{n})+\gamma_{n}(\xi_{n}^{*}\eta_{n}+\eta_{n}^{*}\xi_{n})-
\frac{1}{f}\frac{\partial f}{\partial\omega}S(3\xi_{n}^{*}\xi_{n}+\eta_{n}^{*}\eta_{n})-4\frac{\partial f}{\partial\omega}J\xi_{n}^{*}\xi_{n}\right)d^{d}x
\end{align}
(see \eqref{chargesuper1} and \eqref{Qomega}), which is the sum of the charges of each nonlinear mode, and
\begin{align}\nonumber
&E_{p}=\frac{dE_{0}}{d\omega}\delta\omega+\frac{1}{2}\frac{d^{2}E_{0}}{d\omega^{2}}\delta\omega^{2}+\sum\limits_{j=1}^{d}\frac{E_{0}v_{j}^{2}}{2}
\\ \nonumber&
+\alpha^{2}\omega\sum_{n}\int\left(\omega(\xi_{n}^{*}\xi_{n}+\eta_{n}^{*}\eta_{n})+\gamma_{n}(\xi_{n}^{*}\eta_{n}+\eta_{n}^{*}\xi_{n})-
\frac{1}{f}\frac{\partial f}{\partial\omega}S(3\xi_{n}^{*}\xi_{n}+\eta_{n}^{*}\eta_{n})-4\frac{\partial f}{\partial\omega}J\xi_{n}^{*}\xi_{n}\right)d^{d}x
\\&+
\alpha^{2}\sum_{n}\int \Bigl(\gamma_{n}^{2}(\xi_{n}^{*}\xi_{n}+\eta_{n}^{*}\eta_{n})+\omega\gamma_{n}(\xi_{n}^{*}\eta_{n}+\eta_{n}^{*}\xi_{n})\Bigr)d^{d}x
\end{align}
(see \eqref{energysuper1}, \eqref{QELorj}, and \eqref{Eomega1}), which is the sum of the energies of each nonlinear mode. The obtained results can also be applied to the case of a stable scalar condensate.

As was noted above, the only mode that provides nonzero contributions to $Q_{p}$ and $E_{p}$ that are linear in the corresponding expansion parameter is the mode corresponding to the change of the Q-ball frequency \eqref{modeOmeganonlin}. In this connection, I comment on a possibility of creating a perturbation without changing the total charge of the system. In paper \cite{Smolyakov:2017axd} it was noted that such a process is possible ``if we somehow excite a Q-ball without changing its charge, i.e., if we just add only the energy to the system, it may look like a process changing the original Q-ball plus creating a perturbation'', namely,
\begin{equation}
E_{0}[Q_{0}]\to E_{0}[Q_{0}-Q_{p}]+E_{p}.
\end{equation}
In terms of the modes, this process can be described as creation of the perturbation with a special choice of $\delta\omega$ for mode \eqref{modeOmeganonlin}. Namely, since $\frac{dQ_{0}}{d\omega}\neq 0$ (recall that we consider only classically stable Q-balls with $\frac{dQ_{0}}{d\omega}<0$ ), one can take
\begin{align}\nonumber
&\delta\omega=-\left(\frac{dQ_{0}}{d\omega}\right)^{-1}\\ \label{deltaomega1} &\times\alpha^{2}\sum_{n}\int\left(\omega(\xi_{n}^{*}\xi_{n}+\eta_{n}^{*}\eta_{n})+\gamma_{n}(\xi_{n}^{*}\eta_{n}+\eta_{n}^{*}\xi_{n})-
\frac{1}{f}\frac{\partial f}{\partial\omega}S(3\xi_{n}^{*}\xi_{n}+\eta_{n}^{*}\eta_{n})-4\frac{\partial f}{\partial\omega}J\xi_{n}^{*}\xi_{n}\right)d^{d}x,
\end{align}
leading to $Q_{p}=0$ up to the terms $\sim\alpha^{2}$. For such $\delta\omega$, the perturbation energy takes the form
\begin{equation}\label{EforQzero}
E_{p}^{(Q_{p}=0)}=\sum\limits_{j=1}^{d}\frac{E_{0}v_{j}^{2}}{2}+
\alpha^{2}\sum_{n}\int \Bigl(\gamma_{n}^{2}(\xi_{n}^{*}\xi_{n}+\eta_{n}^{*}\eta_{n})+\omega\gamma_{n}(\xi_{n}^{*}\eta_{n}+\eta_{n}^{*}\xi_{n})\Bigr)d^{d}x,
\end{equation}
where the term $\sim\delta\omega^{2}\sim\alpha^{4}$ is neglected. It was demonstrated in \cite{Smolyakov:2017axd} that the second integral in \eqref{EforQzero} is always positive for nontrivial perturbations, leading to $E_{p}^{(Q_{p}=0)}>0$. Thus, it is clear that an external energy source is necessary to create such a perturbation.

Analogously, a perturbation keeping the total energy of the system is also possible. Indeed, for
\begin{align}\nonumber
&\delta\omega=-\left(\omega\frac{dQ_{0}}{d\omega}\right)^{-1}\times\Biggl(\sum\limits_{j=1}^{d}\frac{E_{0}v_{j}^{2}}{2}+
\alpha^{2}\sum_{n}\int \Bigl(\gamma_{n}^{2}(\xi_{n}^{*}\xi_{n}+\eta_{n}^{*}\eta_{n})+\omega\gamma_{n}(\xi_{n}^{*}\eta_{n}+\eta_{n}^{*}\xi_{n})\Bigr)d^{d}x\\ \label{deltaomega2} &+\alpha^{2}\omega\sum_{n}\int\left(\omega(\xi_{n}^{*}\xi_{n}+\eta_{n}^{*}\eta_{n})+\gamma_{n}(\xi_{n}^{*}\eta_{n}+\eta_{n}^{*}\xi_{n})-
\frac{1}{f}\frac{\partial f}{\partial\omega}S(3\xi_{n}^{*}\xi_{n}+\eta_{n}^{*}\eta_{n})-4\frac{\partial f}{\partial\omega}J\xi_{n}^{*}\xi_{n}\right)d^{d}x\Biggr),
\end{align}
up to the terms $\sim\alpha^{2}$ we get $E_{p}=0$ and
\begin{equation}\label{QforEzero}
Q_{p}^{(E_{p}=0)}=-\frac{1}{\omega}\left(\sum\limits_{j=1}^{d}\frac{E_{0}v_{j}^{2}}{2}+
\alpha^{2}\sum_{n}\int \Bigl(\gamma_{n}^{2}(\xi_{n}^{*}\xi_{n}+\eta_{n}^{*}\eta_{n})+\omega\gamma_{n}(\xi_{n}^{*}\eta_{n}+\eta_{n}^{*}\xi_{n})\Bigr)d^{d}x\right).
\end{equation}
One can see that $\omega Q_{p}^{(E_{p}=0)}<0$. Of course, an external charge source is necessary to create such a perturbation.

Finally, it is interesting to note that the process of Q-ball acceleration, i.e., the process of attaining a constant velocity $\vec v$ such that $|\vec v|\ll 1$, can also be considered as a steplike process of creation of modes \eqref{QELorj}. However, a more consistent description of processes of creation of the modes can be made within the framework of the corresponding quantum theory, which should take into account nonlinearity of the modes. Development of such a theory calls for further investigation.

\section*{Acknowledgements}
The work was supported by Grant No. 16-12-10494 of the Russian Science Foundation.


\begin{thebibliography}{99}
\bibitem{Smolyakov:2017axd}
M.~N.~Smolyakov, Phys.\ Rev.\ D {\bf 97} (2018) 045011 [arXiv:1711.05730 [hep-th]].

\bibitem{Coleman:1985ki}
S.~R.~Coleman, Nucl.\ Phys.\ B {\bf 262} (1985) 263 [Erratum-ibid.\ B {\bf 269} (1986) 744].

\bibitem{Finkelstein:1951zz}
R.~Finkelstein, R.~LeLevier and M.~Ruderman, Phys.\ Rev.\  {\bf 83} (1951) 326.

\bibitem{RosenRosenstock}
N.~Rosen and H.~B.~Rosenstock, Phys.\ Rev.\ {\bf 85} (1952) 257.

\bibitem{Glasko}
V.B.~Glasko, F.~Leriust, Ia.~P.~Terletskii and S.~F.~Shushurin, Sov.\ Phys.\ JETP {\bf 8} (1959) 312.

\bibitem{Rosen0}
G.~Rosen, J.\ Math.\ Phys. {\bf 9} (1968) 996.

\bibitem{Friedberg:1976me}
R.~Friedberg, T.~D.~Lee and A.~Sirlin, Phys.\ Rev.\ D {\bf 13} (1976) 2739.

\bibitem{Kumar:1979sq}
A.~Kumar, V.~P.~Nisichenko and Y.~P.~Rybakov, Int.\ J.\ Theor.\ Phys.\  {\bf 18} (1979) 425.

\bibitem{GSS}
M.~Grillakis, J.~Shatah and W.~Strauss, J.\ Funct.\ Anal.\ {\bf 74} (1987) 160.

\bibitem{Grillakis}
M.~Grillakis, Comm.\ Pure\ Appl.\ Math.\ {\bf 41} (1988) 747.

\bibitem{LeePang}
T.~D.~Lee and Y.~Pang, Phys.\ Rept.\  {\bf 221} (1992) 251.

\bibitem{VK}
N.~G.~Vakhitov and A.~A.~Kolokolov, Radiophys.\ Quantum\ Electron.\ {\bf 16} (1973) 783.

\bibitem{Kolokolov1}
A.~A.~Kolokolov, J.\ Appl.\ Mech.\ Tech.\ Phys.\ {\bf 14} (1973) 426.

\bibitem{Kolokolov2}
A.~A.~Kolokolov, Radiophys.\ Quantum\ Electron. {\bf 17} (1974) 1016.

\bibitem{Panin:2016ooo}
A.~G.~Panin and M.~N.~Smolyakov, Phys.\ Rev.\ D {\bf 95} (2017) 065006 [arXiv:1612.00737 [hep-th]].

\bibitem{Anderson:1970et}
D.~L.~T.~Anderson and G.~H.~Derrick, J.\ Math.\ Phys.\  {\bf 11} (1970) 1336.

\bibitem{MarcVent}
G.~C.~Marques and I.~Ventura, Phys.\ Rev.\ D {\bf 14} (1976) 1056.


\end{thebibliography}
\end{document}